\documentstyle {article}
\font\script=cmmib10
\font\bbold=cmmib10
\newcommand{\SW}{\mbox{\script W}}
\newcommand{\BM}{\mbox{\bbold M}}
\newcommand{\BL}{\mbox{\bbold L}}
\newcommand{\SM}{\mbox{\script M}}
\newcommand{\BR}{\mbox{\bbold R}}
\begin{document}
\textwidth 18.0cm
\textheight 23.0cm
\topmargin -0.5in
\baselineskip 16pt
\parskip 18pt
\parindent 30pt
\title{ \large \bf The averaged null energy condition and difference
inequalities in quantum field theory}
\author{Ulvi Yurtsever \\
{}~~~~~~~~~~~~ \\
Jet Propulsion Laboratory 169-327\\
California Institute of Technology\\
4800 Oak Grove Drive\\
Pasadena, CA 91109\\
and\\
Theoretical Astrophysics 130-33\\
California Institute of Technology\\
Pasadena, CA 91125}
\date{November, 1994\thanks{Submitted to Physical Review D}}
\pagestyle{empty}
\baselineskip 24pt
\maketitle
\thispagestyle{empty}
\vspace{.2in}
\baselineskip 12pt
\begin{abstract}
\noindent For a large class of quantum states, all local (pointwise)
energy conditions widely used in relativity are violated by the renormalized
stress-energy tensor
of a quantum field.
In contrast,
certain nonlocal positivity constraints on the quantum
stress-energy tensor
might hold quite generally, and this possibility has received
considerable attention in recent years. In particular, it is now known
that the averaged null energy condition---the
condition that the null--null component
of the stress-energy tensor integrated along a
complete null geodesic is nonnegative for all states---holds quite
generally in a wide
class of spacetimes for a minimally coupled scalar field. Apart
from the specific class of spacetimes considered (mainly two-dimensional
spacetimes and four-dimensional Minkowski space), the
most significant restriction on this result is
that the null geodesic over which the average is taken must be achronal.
Recently, Larry Ford and Tom Roman have explored this restriction in
two-dimensional flat spacetime, and discovered that in a flat cylindrical
space, although
the stress-energy tensor itself fails to satisfy the averaged
null energy condition (ANEC) along the (non-achronal)
null geodesics,
when the ``Casimir-vacuum" contribution is subtracted from the
stress-energy the resulting tensor does satisfy the ANEC inequality.
Ford and Roman name this class of constraints on the quantum stress-energy
tensor ``difference inequalities." Here I give a proof of the difference
inequality for a minimally coupled massless scalar field in an arbitrary
two-dimensional spacetime, using the same techniques as those we
relied on to prove ANEC in an earlier paper with Robert Wald.
I begin with an overview of averaged energy conditions in quantum field
theory.
\vspace{0.5cm}

\end{abstract}
\newpage
\baselineskip 14pt
\parskip 10pt
\pagestyle{plain}
\pagenumbering{arabic}
{}~~~~~~

{\bf \noindent 1.  Averaged energy conditions}

All general results in relativity require some information about the
matter content of spacetime as input. Although in classical physics the
issue can be avoided by assuming spacetime to be empty, in quantum
theory absolute vacuum is meaningless, and questions about how
matter fields contribute to the semiclassical Einstein equations are
unavoidable whenever quantum effects are important gravitationally.
In classical relativity, information about the matter content
of spacetime can often be specified quite succinctly by the
form of the stress-energy tensor for a specific matter field. For
example, the electromagnetic-field contribution,
\begin{equation}
T_{a b} = \frac{1}{4 \pi} \left( F_{a}^{\; \, c} F_{b c} -
\mbox{${1\over4}$} \,
g_{a b} F^{c d} F_{c d} \right) \; ,
\end{equation}
can be characterized by stating that the stress-energy tensor has the
form Eq.\,(1) for some closed two-form $F_{a b}$ that satisfies the vacuum
Maxwell's equation \mbox{$d \, {\tt \char42} F =0$}. Similar
characterizations exist in principle in quantum field theory. For
example, the regularized stress-energy tensor of a minimally-coupled scalar
field on Minkowski spacetime can be characterized
as any tensor $\langle T_{a b} \rangle$
that has the form of the coincidence limit
\begin{equation}
\langle T_{a b} \rangle (x) =
\lim_{y \rightarrow x} \left[\frac{{\partial}^2}{\partial
x^a \partial y^b} w(x,y) - {1\over2} \, g_{ab} \left( g^{c d}
\frac{{\partial}^2}{\partial x^c \partial y^d}  + m^2
\right) w(x,y) \right]
\end{equation}
in an inertial coordinate system $\{ x^a \}$, where $w(x,y)$ is any
smooth, symmetric bi-solution of the wave equation
such that the two-point function
${\mu}_{0} + w$---where ${\mu}_{0}$ is the symmetric two-point function of
the Poincare vacuum state---satisfies the positivity condition (see
Sect.\,3 below). Already in flat spacetime, a description of the quantum
matter content based on Eq.\,(2) is far more complicated than the
corresponding classical description as illustrated by Eq.\,(1). In
arbitrary curved spacetime, the kind of
characterization of $\langle T_{ab} \rangle$ expressed
in Eq.\,(2), although available in principle, is so heavily cluttered
with formalism as to be essentially intractable.

In fact, even in classical physics a complete, detailed
specification of the
stress-energy as in Eq.\,(1) is seldom very illuminating
unless one's primary
interest is to find exact solutions of Einstein's equations.
Instead, in modern approaches to relativity only the most fundamental
features of $T_{ab}$, distilled from expressions like Eq.\,(1), are
relied on to gain general insight into spacetime
structure. These fundamental features are that
$T_{ab}$ is a conserved, symmetric tensor, and that $T_{ab}$
satisfies certain ``energy conditions" at every point in spacetime;
the energy conditions
express, roughly, the idea that the locally-measured energy density
must be positive everywhere for all observers.
The energy conditions (or, more precisely,
at least the weak energy condition)
are universal in the sense that (i) they are obeyed by
the classical stress-energy tensors of all matter fields, and (ii) they
play a crucial role in deriving most of what we know about
the large-scale structure of spacetime. Indeed, it is
perfectly plausible to regard the energy conditions, along with the
conservation property, as a complete characterization of classical
matter in general relativity.

In deep contrast with this classical picture, in quantum theory no such
compact characterization is yet available for the right hand side of the
semiclassical Einstein equations. The symmetry
and conservation properties still hold
for $\langle T_{ab} \rangle$, of course, but none of the local energy
conditions do. Even in flat, Minkowski space, the regularized
(normal-ordered)
expectation value $\langle \omega | \!\! : \!\! T_{00}(x) \!\! : \!\!
| \omega \rangle$ at any point $x$ is unbounded from below as a
functional of the quantum state $\omega$. One might think that
this is a pathology, as is often the case in quantum field theory,
that stems from localizing the operator $: \!\! T_{00}(x) \!\! :$
to a single point $x$ in spacetime.
It turns out, however, that
the volume integral of
$\langle \omega | \!\! : \!\! T_{00}(x) \!\! : \!\! | \omega \rangle$
over any fixed, spacelike 3-box of {\sl finite} size
is also unbounded from below
as a functional of $\omega$ (and a similar result holds for the
spacetime-volume integral over a compact 4-box). It appears that by
choosing the quantum state $\omega$ appropriately one can stuff
an unbounded amount of negative energy into any fixed, finite region of
spacetime, possibly at the expense of placing more and more positive
energy outside the sharply defined boundaries of that region ([1]).

In the absence of a workable, complete characterization of
regularized stress-energy tensors in quantum field theory,
many of the basic questions about global spacetime
structure in semiclassical gravity remain unanswered. For example,
can spacetime singularities, generically unavoidable with classical
matter, be avoided when quantum effects make the dominant
contribution to stress-energy?  Are classically forbidden configurations
of spacetime curvature (such as traversable
wormholes, certain kinds of topology
change) allowed in semiclassical gravity? Is the total mass of a bounded
lump of quantized matter positive as measured from infinity? More
generally, is any conserved tensor $T_{ab}$ realizable as
the regularized stress-energy tensor of some quantum state? If this were the
case, semiclassical gravity would have almost no physical
content. If it is not the case, then what are the constraints $T_{ab}$
has to satisfy to be a physical stress-energy tensor?
I would like to advocate the
discovery of a useful and complete characterization of
quantum stress-energy tensors as one of
the most important unsolved problems in
curved spacetime quantum field theory.

Promising first steps towards the construction of such a characterization
have appeared in recent years, beginning with [2], and with further
developments in [3]--[8]. An earlier work, Ref.\,[9], discussed
somewhat related issues. The results of these early
investigations concern various nonlocal constraints on the stress-energy
tensor involving
its integrals along causal geodesics;
the most important of
these constraints is the averaged null energy condition (ANEC). In its
simplest form, ANEC constrains the stress-energy tensor $T_{ab}$
such that the integral
\begin{equation}
\int_{\gamma} T_{ab} \, k^a \, k^b \, dv \, \geq \, 0 \; ,
\end{equation}
where $\gamma$ is a complete {\sl null} geodesic with affine
parameter $v$ and corresponding tangent vector $k^a$.
The precise, general formulation of ANEC which does not assume the
convergence of the integral in Eq.\,(3) can be found in Sect.\,2 of
[4]. For a number of significant global results in relativity, ANEC (or at
least a corresponding condition along half-complete null geodesics)
seems to be strong enough to replace the classical pointwise null energy
condition; these include the Penrose singularity theorem ([10]--[12]),
and the positive mass theorem ([13]). For a minimally coupled scalar
field in two dimensions,
it is known that the regularized stress-energy
tensor satisfies ANEC with complete generality,
along all complete achronal null geodesics in any
globally-hyperbolic spacetime, and in every Hadamard
quantum state of the field ([4]). In four dimensions,
this general ANEC result holds in
Minkowski spacetime, and, more generally, in any
spacetime with a bifurcate Killing horizon it holds along the
achronal null generators of the horizon, provided an isometry-invariant
(with respect to the Killing field) Hadamard
state exists (see [4] for details).

Conditions similar to ANEC but with $\gamma$ replaced with a complete
{\sl timelike} geodesic hold with some generality for quantum
stress-energy tensors in Minkowski space ([2], [6], [7]). However,
this appears
to be a special feature of flat-spacetime quantum field theory; it is
not difficult to find curved-spacetime
counterexamples to the timelike averaged weak energy condition ([14]).
In fact, relying on a simple scaling argument,
we pointed out in [4] that even ANEC, although it holds
with complete generality in two dimensions, cannot hold generally in
curved four-dimensional spacetimes (see [16] for a further development
of this scaling idea). Although this argument and its conclusion are
correct, it is by no means clear that they spell the
demise of averaged energy conditions in quantum field theory;
that many people have been led to believe so appears to be the result
of a greatly exaggerated
interpretation of the argument. Elsewhere ([17]) I argue for a more moderate
interpretation, based, essentially, on a generalized version of ANEC in
which the right hand side of Eq.\,(3) is replaced with a more general finite
lower bound. I do regard ANEC-type inequalities as a very promising
starting point towards the formulation of
a complete set of universal constraints
that characterize quantum stress-energy tensors. For example,
one might think that ANEC is too weak because it only constrains
the integrals of $T_{ab}$ along null geodesics;
in fact, ANEC appears to be powerful enough to place constraints on other,
more general averages of the stress-energy tensor. I will now present a
simple analysis that supports this view:

I will consider a stress-energy tensor $T_{ab}$ that satisfies the simple
form of ANEC given by Eq.\,(3);
therefore I implicitly assume that $T_{ab}$ falls off appropriately
at infinity. A more sophisticated analysis that does away with this
assumption (and possibly with some of the other strong assumptions I
will make below) can probably be given; I retain these assumptions to keep
my analysis transparent. More precisely, I consider
a globally hyperbolic spacetime $(\mbox{\script M}, g)$, and
a conserved stress-energy tensor $T_{ab}$ which satisfies
ANEC in the form of Eq.\,(3) along all complete, achronal null geodesics in
$(\mbox{\script M}, g)$. Let $\Sigma \subset \mbox{\script M}$
be a spacelike Cauchy surface, and
$S \subset \Sigma$ be a compact submanifold (with
boundary) in $\Sigma$.
\vspace{5pt} I assume:
\newline (A1) The subregion $S \subset \Sigma$
is chosen large enough such that
ANEC [Eq.\,(3)] holds along the null generators of the future horizon
$H^{+}(S)$\vspace{5pt}.
\newline Note that the generators of $H^{+}(S)$ are necessarily
achronal null geodesics; they have their past endpoints on the boundary
$\partial S$ of $S$, and their future endpoints on a
(in general complicated) caustic set $\mbox{\script C}$. I assume
\vspace{5pt} that:
\newline (A2) There exists a time function $\alpha$
defined on the domain of dependence $D^{+}(S)$ such that
(i) $\alpha = 0$ on $S$ and $\alpha = 1$ on $H^{+}(S)$,
(ii) $\alpha_{;a}=- \kappa \, n_{a}$ on $S$,
where $n^{a}$ is the future-pointing
unit normal to $S$ and $\kappa$ is a positive constant, and (iii)
throughout the interior of $D^{+}(S)$
\begin{equation}
T^{ab} \alpha_{;ab} \, \leq \, q \, T^{ab} \alpha_{;a} \alpha_{;b}
\end{equation}
for some constant \vspace{5pt}$q > 0$.
\newline{\bf Theorem}: Under the assumptions (A1) and (A2), the total
energy contained in the region $S \subset \Sigma$ is nonnegative:
\begin{equation}
\int_{S} T^{ab}n_{a}n_{b} \, d^3 \sigma \, \geq \, 0 \;,
\end{equation}
where $d^3 \sigma$ is the volume element on $\Sigma$.

The assumption (A1) is quite reasonable within the scope of the present
discussion; only the last part of assumption (A2) [i.e., the inequality
Eq.\,(4)] is unpleasantly strong.
(See Fig.\,1 for a two-dimensional representation of the geometry
involved in this analysis).
Notice that $T_{ab}$ does not
necessarily satisfy the (pointwise) weak energy condition,
so the right hand side of Eq.\,(4) is not necessarily positive, and
simply choosing a large $q > 0$ will not do. Moreover, because
of its construction the time function
$\alpha$ has to develop gradient singularities at the boundary $\partial S$
and at the caustic set $\mbox{\script C}$ [but $\alpha$ is smooth
throughout the (open) interior of the domain of dependence $D^{+}(S)$],
so Eq.\,(4) effectively constrains the asymptotic behavior
of $T_{ab}$ near $\mbox{\script C}$ and $\partial S$ (see Fig.\,1).
For a typical example of the geometry involved here consider
two-dimensional Minkowski spacetime, with the surface $\Sigma$ given by
$\{ t=-K \}$, $K > 0$, and with $S \subset \Sigma$ given by that piece of
$\Sigma$ lying within the wedge $\{ |x| \leq |t| \}$. In this
example the horizon
$H^{+}(S)$ is the past null cone of the origin, $H^{+}(S)=\{
|t|=|x|, \; -K \leq t \leq 0 \}$, and for $|x| \ll |t|$ the time
function $\alpha$ can be taken as
\[
\alpha = \frac{1}{K} \left( K + t - \frac{x^2}{t} \right)
\]
up to corrections of order $(x^2/t^2)$. This function satisfies
the conditions (i) and (ii) of (A2) [again up to order ($x^2/t^2$)]
with $\kappa=1/K$ . Then, assuming that $T_{00}$ is the
dominant component (in absolute value) of $T_{ab}$,
\begin{eqnarray}
T^{ab} \alpha_{;ab} & \approx & T^{00} \, \alpha_{,tt} = - T^{00} \;
\frac{2}{K t} \frac{x^2}{t^2} \; , \nonumber \\
T^{ab} \alpha_{;a} \alpha_{;b} & \approx & T^{00} \, \alpha_{,t}^2
= T^{00} \; \frac{1}{K^2} \left( 1 + \frac{x^2}{t^2} \right) \; . \nonumber
\end{eqnarray}
Therefore, in the region $|x| \ll |t|$ throughout the domain of dependence
$D^{+}(S)$, the
quantity $T^{ab} \alpha_{;ab}$ has the same sign as $T^{ab}\alpha_{;a}
\alpha_{;b}$ and is down in magnitude
by a factor $x^2/t^2$. With the provision that
$T^{ab}$ falls off nicely at the boundary $\partial S$ and near the caustic
set (here the origin)
$\mbox{\script C}$ (where the true $\alpha$ becomes singular), and with
the constant $q$ chosen in the range $0 < q < 1$,
this Minkowski-spacetime example suggests
the inequality Eq.\,(4) to be a reasonable assumption.
Quite possibly this inequality
can be weakened considerably without changing the main
argument in the proof of the theorem, which I now proceed to
\vspace{5pt}give.
\newline {\bf Proof of the theorem}: Let me define (see Fig.\,1)
\begin{eqnarray}
S_{u} & \equiv & \{ p \in D^{+}(S) \; | \; \alpha(p) = u \} \; ,
\nonumber \\
V_{u} & \equiv & \{ p \in D^{+}(S) \; | \; \alpha(p) \leq u \} \; .
\nonumber
\end{eqnarray}
By integrating the identity
\[
(T^{ab} \alpha_{;a})_{;b} = T^{ab} \alpha_{;ab}
\]
over the volume $V_{u}$, I obtain
\begin{equation}
- \int_{S_{u}} T^{ab} \alpha_{;a} n_{b} \, d^3 \sigma
\, + \, \int_{S_{0}} T^{ab} \alpha_{;a} n_{b} \, d^3 \sigma
\, = \, \int_{V_{u}} T^{ab} \alpha_{;ab} \, dV \; ,
\end{equation}
where $n_b$ denotes the future-pointing unit normal and
$d^3 \sigma$ is the volume form on $S_{u}$. Putting
\begin{equation}
I(u) \equiv - \int_{S_{u}} T^{ab} \alpha_{;a} n_{b} \, d^3 \sigma \; ,
\end{equation}
and combining Eq.\,(6) with Eq.\,(4), I obtain the inequality
\begin{equation}
I(u) \leq I(0) + q \int_{V_{u}} T^{ab} \alpha_{;a} \alpha_{;b} \, dV \;.
\end{equation}
Since $\alpha_{;b}$ is parallel and opposite to the
future-pointing unit normal $n_b$,
the second term on the right hand side of Eq.\,(8) can be written as
\begin{eqnarray}
\int_{V_{u}}T^{ab}\alpha_{;a}\alpha_{;b} \, dV & = &
- \int_{0}^{u} d \alpha \, \int_{S_{\alpha}} T^{ab} \alpha_{;a} n_b \,
d^3 \sigma \nonumber \\
& = & \int_{0}^{u} d \alpha \, I(\alpha ) \; ;
\end{eqnarray}
hence the inequality Eq.\,(8) takes the form
\begin{equation}
I(u) \leq I(0) + q \int_{0}^{u} I(\alpha ) \, d \alpha \; .
\end{equation}
Now inspect the definition Eq.\,(7) of the quantity $I(u)$. On the
null surface \mbox{$\{ \alpha \! = \! 1 \}  =
H^{+}(S)\,$}, $\; \alpha^{;a}$ (since
it is a null gradient) is necessarily
an affine tangent vector along the null geodesic generators.
Therefore, $I(1)$ is the average over the set of null generators
of $H^{+}(S)$ of the ANEC integrals of $T_{ab}$
along those generators. Thus $I(1) \geq 0$ by assumption
(A1). But it is clear from the inequality Eq.\,(10) that the
conclusion $I(1) \geq 0$ is incompatible with the assumption $I(0) < 0$.
Therefore I conclude that $I(0) \geq 0$. By assumption (A2)-(i) and
(A2)-(ii)
\[
I(0) = \kappa \int_{S} T^{ab} n_a n_b \, d^3 \sigma \; ,
\]
and the assertion
of the theorem [Eq.\,(5)] follows.\raisebox{-.3ex}{$\;\Large\Box$}

{}~~~~~~~~

{\bf \noindent 2. Difference inequalities}

As I mentioned briefly in the previous section,
ANEC needs to be generalized since in the strict form given by Eq.\,(3)
it is typically violated in four-dimensional curved
spacetimes ([4], [16]). The most natural generalization of ANEC
involves replacing the right hand side of Eq.\,(3) with a broadly
specified lower bound; I will discuss this in more detail in [17].
Another, closely related generalization
has been discovered by L.\ Ford and T.\
Roman in [18]; it involves what they term ``difference inequalities."
A difference inequality is an ANEC-type inequality of the form
\begin{equation}
\int_{\gamma} \left( \, \langle
\omega | T_{ab}
| \omega \rangle \, - \, D_{ab} \, \right)\,k^{a}k^{b} \, dv \;
\geq \; 0 \; \; \; \; \; \; \; \; \forall \omega \; ,
\end{equation}
where $\langle \omega | T_{ab} | \omega \rangle$ denotes the
renormalized stress-energy tensor in the quantum state $\omega$,
$D_{ab}$ is a state-independent, geometric tensor on spacetime, and the
integral is evaluated along a complete
null geodesic $\gamma$ as in Eq.\,(3).
The difference inequality can be given a more
precise formulation that does not require the convergence of the
integral in Eq.\,(11) in just the same way as ANEC
[see Sect.\,2 of [4] and Eq.\,(15) below].
If the integral
\begin{equation}
\int_{\gamma} D_{ab} \, k^{a}k^{b} \, dv
\end{equation}
converges, Eq.\,(11) yields a (in general nonzero)
lower bound on the ANEC integral:
\begin{equation}
\int_{\gamma}  \langle
\omega | T_{ab}
| \omega \rangle \, k^{a}k^{b} \, dv \; \geq \;
\int_{\gamma} D_{ab} \, k^{a}k^{b} \, dv \; \; \; \; \; \; \; \;
\forall \omega \; .
\end{equation}
It is this form of the difference inequality which makes it potentially
significant for applications such as singularity theorems (see [17]).

What Ford and Roman have discovered in [18] is that
for a massless Klein-Gordon field on the flat cylinder (two-dimensional
Minkowski spacetime identified modulo a discrete group of
spatial translations),
the difference inequality Eq.\,(11) holds along
all complete null geodesics
provided $D_{ab}$ is the stress-energy tensor of
the Casimir vacuum state:
\begin{equation}
D=\frac{\pi}{6 L^2} \left[ \begin{array}{cc}
           -1 & 0 \\
            0 & -1
         \end{array}      \right] \; ,
\end{equation}
where $L$ is the length of the spatial sections (i.e., the points $(x,t)$
and $(x+L,t)$ in Minkowski spacetime are identified). Note that complete
null geodesics on the cylinder are {\sl not} achronal, and ANEC is violated
along them [e.g.\ in the Casimir vacuum state
where $\langle T_{ab} \rangle$
is given by the right hand side of Eq.\,(14)]. Also note that the
difference inequality cannot have the form Eq.\,(13)
in this case since the integral Eq.\,(12) does not converge.
Nevertheless, this two-dimensional difference-inequality
result of [18] relaxes the achronality assumption of the ANEC theorem
proved in [4], and its main significance lies in
this improvement.

In the remaining sections of the paper I will give a general proof of
the difference inequality for a massless Klein-Gordon field in
an arbitrary, globally hyperbolic two-dimensional curved spacetime
$(\mbox{\script M},g)$. More
precisely, for every Hadamard state $\omega$ of the field and for every
complete null geodesic $\gamma$ in $(\mbox{\script M},g)$ with
affine parameter $v \in (-\infty , \infty)$,
I will prove that
the following holds: Let $c(x)$ be any
bounded real-valued function of compact support on $\mbox{\bbold R}$
whose Fourier
transform $\hat{c} (k)$ is such that
for some $\delta > 0$ the function
$(1+ k^{2})^{1 + \delta } | \hat{c} (k) |$
is bounded
[i.e.,
$| \hat{c} (k) |$ decays at least as fast as $|k|^{-2 - 2 \delta }$ as
$|k| \rightarrow \infty $; this implies that $c(x)$ is $C^{1}$.]
Then, for all choices of origin of the affine parameter $v$,
the regularized stress-energy tensor
$\langle \omega | T_{a b} | \omega \rangle$
satisfies the inequality
\begin{equation}
\liminf_{ \lambda \rightarrow \infty } \int_{- \infty }^{ \infty }
(\, \langle \omega | T_{a b} | \omega \rangle \, - \, D_{ab}
\, ) \, k^{a} k^{b} \,
[c( v/ \lambda )]^{2} \, dv \, \geq \, 0 \; ,
\end{equation}
where $D_{ab}$ is a
state-independent tensor which I will specify
[$D_{ab}$ depends only on the geometry of
$(\mbox{\script M},g)$]. When the integrand
$(\langle \omega | T_{a b} | \omega \rangle - D_{ab}
) \, k^{a} k^{b}$ is integrable, Eq.\,(15) reduces to the simple form
Eq.\,(11) of the difference inequality.

My proof of the difference inequality Eq.\,(15) will be entirely
parallel to the proof of ANEC in two dimensions that
we gave in Sects.\,4 and 5
of [4], and I will omit all details which are simply
restatements of the corresponding details in [4] modified to suit the
present analysis. Consequently, readers who wish to follow the remaining
sections of this paper closely will find it useful to have a copy of [4]
at hand while they do so. I will begin, in Sect.\,3 below, by
discussing the relationship between quantum states on
\mbox{Minkowski spacetime
$\mbox{\bbold R}^2$} and quantum states on the flat cylinder
$S^1 \times \mbox{\bbold R}$. I describe the proof of the
difference inequality for the flat cylinder
$S^1 \times \mbox{\bbold R}$ in Sect.\,4, and for a
general, curved two-dimensional spacetime in Sect.\,5.

{}~~~~~~~

{\bf \noindent 3. States on $\mbox{\bbold R}^2$ and states on $S^1
\times \mbox{\bbold R}$}

As we did throughout [4], so also here I will adopt the algebraic
viewpoint on quantum field theory in a curved
(globally hyperbolic) spacetime $(\mbox{\script M}, g)$. In particular,
quantum states $\omega$ are specified by their two-point distributions
\begin{eqnarray}
\lambda [F,G] & \equiv & \omega ( \phi [F] \phi [G] ) \; ,
\; \; \; \; \; \; \; \; F, \; G
\in S(\mbox{\script M}) \; ,\\
\lambda (f,g) & \equiv & \lambda [Ef, Eg] \; ,
\; \; \; \; \; \; \; \; \; \; \; \; f, \; g \in
 C_{0}^{\infty}(\mbox{\script M}) \; ,
\end{eqnarray}
where $S(\mbox{\script M})$ is the space of all
solutions of the Klein-Gordon equation
which are compact supported on Cauchy surfaces,
and $Ef \in S(\mbox{\script M}) $ denotes the
``advanced minus retarded" solution with source
$f \in C_{0}^{\infty}(\mbox{\script M})$ (see Sect.\,3 of [4] for a
concise introduction to the algebraic approach).
The two-point function $\lambda$ can be written in the form
\begin{equation}
\lambda = \mu + \mbox{$1\over 2$} i \, \sigma \; ,
\end{equation}
where $\mu [F,G] = {\rm Re} (\lambda [F,G])$, and $\sigma [F,G]$ is the
Klein-Gordon inner product (see Sect.\,3 in [4] for details).
The algebraic positivity
condition on the state $\omega$ is equivalent to the inequalities
$\mu [F,F] \geq 0 \; \; \forall F \in S(\mbox{\script M})$, and
\begin{equation}
\mu [F,F] \, \mu [G,G] \, \geq \, \mbox{$1 \over 4$} | \sigma [F,G] |^2
\; \; \; \; \; \; \; \; \forall F, \; G \in S(\mbox{\script M}) \; .
\end{equation}
Note that Eq.\,(19) implies $\mu [F,F] > 0$ for all $F \neq 0$
in $S(\mbox{\script M})$.

Let $\mbox{\bbold M}$ denote the two-dimensional
Minkowski spacetime $(\mbox{\bbold R}^2, \eta)$, where $\eta = dx^2 -
dt^2 = -du \, dv$, and let $\mbox{\bbold L}$ denote the flat
cylinder $S^1 \times \mbox{\bbold R}$ obtained from Minkowski space
by the identification $(x,t) \equiv (x+L,t)$. There exists a canonical
``wrapping" map $\mbox{\script W} : C_{0}^{\infty}(\mbox{\bbold M})
\rightarrow
C_{0}^{\infty}(\mbox{\bbold L})$ given by
\begin{equation}
\mbox{\script W} : f(x,t) \; \longmapsto
\sum_{n=-\infty}^{\infty} f(x+nL,t) \; ,
\; \; \; \; \; \; \; \; f \in C_{0}^{\infty}(\mbox{\bbold M}) \; ,
\end{equation}
and a corresponding ``wrapping"
map
$\mbox{\script W}_{S} : S(\mbox{\bbold M})
\rightarrow
S(\mbox{\bbold L})$ which satisfies \mbox{$\mbox{\script W}_{S} \circ
E_{M} = E_{L} \circ \mbox{\script W}$}, where $E_{M}$ and $E_{L}$ denote
the advanced-minus-retarded Green's functions on $\mbox{\bbold M}$ and
$\mbox{\bbold L}$, respectively. I will denote $\mbox{\script W}_{S}$ by
the same symbol as $\mbox{\script W}$ as long as
it is clear from the context which map is which. Also, with
$f \in C_{0}^{\infty}(\mbox{\bbold M})$ and $F \in S(\mbox{\bbold M})$,
I will use the shorthand notation $f^W$ and $F^W$ to denote
$\mbox{\script W} (f) \in C_{0}^{\infty}(\mbox{\bbold L})$ and
$\mbox{\script W} (F) \in S(\mbox{\bbold L})$, respectively.
Note that the map $\SW$ is onto both from $C_{0}^{\infty}(\BM )$ to
$C_{0}^{\infty}(\BL )$ and from $S(\BM )$ to $S(\BL )$.

Let $\phi$ be a massless Klein-Gordon field on $\BM$, and similarly on
$\BL$. A Hadamard ([4]) quantum state $\omega$ on $\BM$ is specified by a
two-point function of the form
\begin{equation}
\mu_{M}(x,x') = \mu_{0M} (x,x') + w_{M}(x,x') \; ,
\end{equation}
where $\mu_{0M}$ is the two-point function of the Poincare vacuum
[given by Eq.\,(19) in [4], Sect.\,4],
and $w_M$ is a smooth, symmetric
bi-solution of the Klein-Gordon equation such that
$\mu_M=\mu_{0M}+w_{M}$ satisfies the positivity inequality
\begin{equation}
\mu_{M}(f,f) \, \mu_{M}(g,g) \, \geq \, \mbox{$1 \over 4$}
| \sigma_{M} (f,g) |^2 \; \; \; \; \; \; \; \; \forall
f, \; g \in C_{0}^{\infty}(\BM ) \;.
\end{equation}
On the flat cylinder $\BL$, the analogue of the Poincare vacuum
state is the Casimir vacuum, which can be constructed, e.g., by a mode
decomposition where positive-frequency solutions are
defined with respect to the canonical timelike Killing vector on $\BL$.
The Casimir vacuum is a Hadamard state and I will denote its two-point
function by $\mu_{0L}$ [see Eq.\,(28) below].
The two-point function of any other Hadamard
state on $\BL$ can be written in the form
\begin{equation}
\mu_{L}(x,x') = \mu_{0L} (x,x') + w_{L}(x,x') \; ,
\end{equation}
where $w_{L}$ is a smooth, symmetric
bi-solution of the Klein-Gordon equation on
$\BL$ such that $\mu_{L}$ satisfies the positivity
inequality appropriate for $\BL$:
\begin{equation}
\mu_{L}(f,f) \, \mu_{L}(g,g) \, \geq \, \mbox{$1 \over 4$}
| \sigma_{L}(f,g) |^2 \; \; \; \; \; \; \; \; \forall
f, \; g \in C_{0}^{\infty}(\BL ) \;.
\end{equation}
What is the relationship between Hadamard states
on $\BL$ and Hadamard states on the
Minkowski spacetime $\BM$? To explore this question, it is convenient to
pull back distributions defined on $\BL$ (such as $\mu_{L}$ and
$\sigma_{L}$) via the wrapping map $\SW$ so that they become
distributions on $\BM$. For example, the pull-back
of the distribution $\mu_{L}$ is the
distribution ${\mu_{L}}^M$ on $\BM$
defined by
\begin{equation}
{\mu_{L}}^{M} (f,g) \, \equiv \, \mu_{L} (f^W,g^W) \; ,
\; \; \; \; \; \; \; \;  f, \; g \in C_{0}^{\infty}(\BM ) \; ,
\end{equation}
and the pull-backs ${\mu_{0L}}^{M}, \;
{w_{L}}^{M}$ and ${\sigma_{L}}^{M}$
of the distributions $\mu_{0L}, \; w_{L}$
and $\sigma_{L}$ are defined similarly. Since the wrapping map
$\SW$ is onto, I can now rewrite the
positivity condition for $w_{L}$ [Eq.\,(24)] as an inequality defined
purely on $\BM$:
\begin{eqnarray}
({\mu_{0L}}^M + {w_L}^M )(f,f) \;
({\mu_{0L}}^M + {w_L}^M )(g,g) & \geq & \mbox{$1 \over 4$}
| {\sigma_{L}}^M (f,g) |^2 \nonumber \\
& \forall &
f, \; g \in C_{0}^{\infty}(\BM ) \;.
\end{eqnarray}
Is ${\mu_{L}}^M={\mu_{0L}}^M + {w_L}^M$
the two-point function of a Hadamard state on Minkowski spacetime?
Clearly, ${\mu_{L}}^M$ in general, and ${\mu_{0L}}^M$ in particular, all
have the correct short distance behavior to be genuine Hadamard states
on $\BM$ ([19]). However, it is easy to see ([20]) that ${\mu_{L}}^M$ fails
to satisfy the positivity condition Eq.\,(22) on $\BM$, hence it fails
to be a quantum state to begin with. The reason is simple: there exist
many nonzero $f \in C_{0}^{\infty}(\BM )$ such that $E_M f \neq 0$ and
$\SW (f) = 0$; in other words, there exist nonzero $F \in S(\BM )$ such
that $F^W =0$. Thus, for each such $F \in S(\BM )$, $F \neq 0$ and
${\mu_{L}}^M [F,F] = 0$, and this contradicts the positivity inequality
on $\BM$.

The next natural question to ask is whether ${w_L}^M$ is the regularized
two-point function of a Hadamard state on $\BM$. In other words, given a
smooth, symmetric
bi-solution ${w_L}$ on $\BL$ which satisfies the inequality
Eq.\,(26), is the two-point function $\mu_{0M} + {w_{L}}^M$ a Hadamard
state on Minkowski spacetime? This two-point function has
the Hadamard form by definition, and, physically, would make sense as the
``unwrapping" of the quantum state ${\mu}_{0L} + w_L$ onto the
covering space $\BM$. Clearly, the argument above proving
${\mu_{0L}}^M+{w_L}^M$ to be in violation of positivity does not apply
to ${\mu}_{0M} + {w_L}^M$. Furthermore, if ${\mu}_{0M}+{w_L}^M$ were
indeed a state on $\BM$, the proof of the difference inequality
Eq.\,(15) on $\BL$ would follow immediately from the proof of ANEC
on $\BM$: the regularized stress-energy of this ``state" on
Minkowski spacetime is precisely the difference $\langle T_{ab}
\rangle_{L} - D_{ab} $ which appears in the integrand in Eq.\,(15).
Unfortunately (for my purposes), the two-point function
$\mu_{0M}+{w_L}^M$ in general
does violate the positivity condition on Minkowski
space, and therefore does not represent a Minkowski quantum state
for every choice of $w_L$. The
demonstration of this is only slightly more involved than the
argument I gave in the preceding paragraph (which proved that
${\mu_{0L}}^M+{w_L}^M$ violates positivity). Namely, consider the
following expressions for the two-point distributions $\mu_{0M}$ and
${\mu_{0L}}^M$:
\begin{eqnarray}
\mu_{0M}(f,f) & = & \int_{-\infty}^{\infty}
\frac{\pi}{|k|} \, dk \, | \hat{f}(k,-|k|)|^2 \; , \\
{\mu_{0L}}^M (f,f) & = & {\sum_{n=-\infty}^{\infty}} {\! \!}' \;
\frac{\pi}{|k_{n}|}
\, (\Delta k) \, | \hat{f}(k_n , -|k_n |) |^2 \;
, \; \; \; \; \; \; \; f \in C_{0}^{\infty}(\BM ) \; ,
\end{eqnarray}
where $\hat{f}(k , \omega )$ denotes the Fourier transform
\[
\hat{f}(k, \omega ) \equiv \frac{1}{2 \pi} \int_{-\infty}^{\infty}
\int_{-\infty}^{\infty} e^{ - i (kx + \omega t) } f(x,t) \, dx \, dt
\]
of $f \in C_{0}^{\infty}(\BM )$, $k_n = 2 \pi n /L$, and $\Delta k
\equiv k_{n+1}-k_n = 2 \pi / L$.
The prime on the summation sign in Eq.\,(28)
indicates that the sum excludes $n=0$; this is because massless quantum
field theory in two dimensions is formulated with test functions $f
\in C_{0}^{\infty}(\SM )$ which satisfy $\int f = 0$ to avoid
infrared-divergence problems. [See the paragraph following Eq.\,(17) in
[4] for an explanation of this. Note that I ignore this complication
almost completely throughout my analysis in this paper because
accommodating it would
not make any difference (except for making
my notation more complicated than
it already is) in the flow of my argument.]
Now the bi-solution $w_L$ is
required to satisfy only the inequality Eq.\,(26), so that
\begin{equation}
( {\mu_{0L}}^M + {w_L}^M )(f,f) \, \geq \, 0 \; \; \; \; \; \; \; \;
\forall f \in C_{0}^{\infty}(\BM ) \; ,
\end{equation}
whereas to be the regularized two-point function of
a quantum state on Minkowski space it would need to
satisfy
\begin{equation}
( \mu_{0M} + {w_L}^M ) (f,f) \, \geq \, 0 \; \; \; \; \; \; \; \;
\forall f \in C_{0}^{\infty} (\BM ) \; .
\end{equation}
It is therefore sufficient to show that there exist $w_L$ which satisfy
Eq.\,(29) but violate Eq.\,(30). [Note that there exist many $w_L$ which
satisfy both Eqs.\,(29) and (30); any $w_L$ which, as a
bi-solution, decomposes into the tensor product of a solution with itself
is an example of this. In fact, for such a two-particle state on $\BL$,
the ``unwrapping" $\mu_{0M}+{w_L}^M$ {\sl does} correspond to a
genuine state on
Minkowski spacetime.] To see that this is indeed the case, consider
those $f \in C_{0}^{\infty}(\BM )$ whose Fourier transforms
$\hat{f}(k, \omega )$ are sharply
peaked around the quantized frequencies
$k=k_n, \; \omega = -|k_n|$. [To find
a compact supported
$f$ of this kind, start with a $\hat{f}(k, \omega )$
in $L^2(\BR^2 )$ which is
such that (i) for fixed $\omega_0$,
$\hat{f}(k,\omega_0 )$ decays at infinity
faster than any inverse polynomial and is the
restriction to $\BR$ of an entire function $\hat{f_1}(z)$ on
$\mbox{\bbold C}$, and, similarly,
for fixed $k_0$, $\hat{f}(k_0,\omega )$ decays
at infinity faster than any inverse polynomial and is
the restriction to $\BR$ of an entire function $\hat{f_2}(z)$ on
$\mbox{\bbold C}$,
(ii) at complex infinity, $\hat{f_1}(z)$ has the asymptotic behavior
$|\hat{f_1}(z)| \leq M_1 e^{a_1|z|}$, and, similarly,
$\hat{f_2}(z)$
has the asymptotic behavior
$|\hat{f_2}(z)| \leq M_2 e^{a_2|z|}$, $a_i > 0, \; M_{i} > 0$,
and (iii) $\hat{f}(k,\omega )$ is
sharply peaked around $(k_n ,-|k_n |)$,
$n \in \mbox{\bbold Z}$. The inverse Fourier
transform $f(x,t)$ is then guaranteed to be $C^{\infty}$ and of compact
support by Theorem 7.\,4 in Chapter 6 of [21].]
According to Eqs.\,(27)--(28), when smeared with these $f$,
${\mu_{0L}}^M (f,f)$ becomes
arbitrarily large in comparison with
$\mu_{0M} (f,f)$ as
$\hat{f}(k, \omega )$ get more and more sharply
peaked around $(k_n, -|k_n |)$. It is then clear that
a $w_L$ can be found such that
\mbox{${\mu_{0L}}^M + {w_L}^M$} satisfies Eq.\,(29)
for all $f \in C_{0}^{\infty}(\BM )$, but
for these ``spiky"
$f$ peaked at $(k_n , -|k_n |)$, $\, {w_L}^M (f,f)$
manages to be more negative than
$- \mu_{0M}(f,f)$ and hence violates Eq.\,(30).

Although the obvious
``easy" proof of the difference inequality Eq.\,(15)
seems to be ruled out by the above
argument, the results of this section already
provide all the necessary extra ingredients which, when combined with
the proof of ANEC in Sect.\,4 of [4], constitute a complete proof of
Eq.\,(15) as I will now explain.

{}~~~~~~~

{\bf \noindent 4. Proof of the difference inequality in flat
$S^1 \times \mbox{\bbold R}$ spacetime}

As this section follows Sect.\,4 of [4] very closely, for
brevity I will use the
prefix 4 to denote equations in [4]; e.g., Eq.\,(4.\,30) will denote
Eq.\,(30) of Reference [4]. Let $\mu_{0L} + w_L$ be a
Hadamard state on $\BL$ of the form Eq.\,(23), and
let $\gamma$ be a complete null geodesic in
$\BL $. Just as the two-point distributions on $\BL $ were pulled back
to $\BM$ in the previous section, so can the null geodesic $\gamma$ be
lifted to a complete null geodesic $\tilde{\gamma}$ on $\BM$, I can
then carry out my analysis entirely on Minkowski spacetime as in
Sect.\,4 of [4]. Assume, without loss of generality, that
$\tilde{\gamma}= \{ u=0 \}$. Define
\begin{equation}
Y_L (v,v') \equiv \frac{{\partial}^{2}}{\partial v \partial v'}
{w_L}^M (0,v,0,v') \; .
\end{equation}
It is then obvious that along $\tilde{\gamma}$,
\begin{equation}
( \, \langle T_{a b} (v) \rangle_L \,
- \, D_{ab} \, )  k^{a} k^{b} =
(\, \langle T_{vv} \rangle_L \, - \, D_{vv} \, ) =
Y_L (v,v) \; ,
\end{equation}
where $\langle T_{ab} (v) \rangle_L$ denotes the regularized stress-energy
along $\gamma$ pulled back to $\tilde{\gamma}$, and $D_{ab}$ is the
stress-energy tensor [Eq.\,(14)] of the Casimir vacuum state
$\mu_{0L}$, again pulled back to
$\tilde{\gamma}$. In
precise but cumbersome notation (which therefore I will avoid) these
quantities really should be written as $\pi^{\ast}(\langle
T_{ab} \rangle_L )$ and $\pi^{\ast} D_{ab}$, where $\pi : \BM \rightarrow
\BL$ is the canonical projection. Now, by applying exactly the same
arguments as those in [4] leading to Eq.\,(4.\,30), I deduce from the
positivity inequality Eq.\,(26) that for all functions $F_1 , \;
F_2 \in C_{0}^{\infty} ( \BR )$,
\begin{eqnarray}
\lefteqn{ \left[ 2 {{\mu}_{0L}}^M
[ F_{1} , F_{1} ] +  8 \int_{- \infty}^{\infty}
\int_{- \infty}^{\infty} Y_L
(v,v') F_{1} (v) F_{1} (v') \, dv \, dv' \right]
\times  \nonumber } \\
& \times & \left[ 2 {{\mu}_{0L}}^M
[F_{2} , F_{2} ] + 8 \int_{- \infty}^{\infty}
\int_{- \infty}^{\infty} Y_L
(v,v') F_{2} (v) F_{2} (v') \, dv \, dv' \right]
\nonumber \\
&  \geq  & | {\sigma_L}^M [ F_{1} , F_{2} ] |^{2} \; ,
\end{eqnarray}
where now instead of Eqs.\,(4.\,26) and (4.\,31) I have, in accordance
with Eq.\,(28),
\begin{equation}
{{\mu}_{0L}}^M [F,F] =
2 \sum_{n=0}^{\infty}
k_n | \hat{F} (k_n ) |^{2} \, \Delta k \; ,
\end{equation}
and
\begin{equation}
{\sigma_L}^M [F_1 , F_2 ] =
-4 \, {\rm Im} \, \sum_{n=0}^{\infty} k_n
\overline{\hat{F_1} (k_n ) } \hat{F_2} (k_n ) \, \Delta k \;.
\end{equation}
I will now trace the arguments in [4] following Eq.\,(4.\,31) and verify
that they lead to a proof of the difference inequality Eq.\,(15) as
promised. First assume, as in [4], that the function $Y_L (v,v')$
belongs to Schwartz space, i.e., it and all its derivatives decay at
infinity faster than any polynomial. By the same algebra that
leads in [4] to Eq.\,(4.\,32), it follows from Eqs.\,(33)--(35) that
for all $F_1 , \; F_2 \in C_{0}^{\infty}( \BR )$
\begin{eqnarray}
\lefteqn{ \left[ \sum_{n=0}^{\infty} k_n | \hat{F_{1}} (k_n )
|^{2} \, \Delta k
+ \xi ( \hat{F_{1}} , \hat{F_{1}} ) - \eta (
\hat{F_{1}} , \hat{F_{1}} ) \right] \times \nonumber } \\
& & \times \left[ \sum_{n=0}^{\infty} k_n | \hat{F_{2}} (k_n )
|^{2} \, \Delta k
+ \xi ( \hat{F_{2}} , \hat{F_{2}} ) -  \eta (
\hat{F_{2}} , \hat{F_{2}} ) \right] \nonumber \\
& & \geq
\left[ {\rm Im} \sum_{n=0}^{\infty} k_n \overline{\hat{F_{1}} (k_n ) }
\hat{F_{2}} (k_n ) \, \Delta k \right]^{2} \; ,
\end{eqnarray}
where $\xi (\hat{F},\hat{F})$ and $\eta (\hat{F},\hat{F})$ are given by
the same expressions as Eqs.\,(4.\,33) (with $\hat{Y}$ replaced with
$\hat{Y}_L$). Then, precisely the same argument which in [4] leads from
Eq.\,(4.\,32) to Eqs.\,(4.\,39) and (4.\,40) leads here to the
conclusions
\begin{equation}
\hat{Y}_L ( \kappa , - \kappa ) \geq 0 \; \; \; \; \; \; \forall \kappa
\in [0, \infty) \; ,
\end{equation}
and, as in Eq.\,(4.\,40),
\begin{equation}
\int_{- \infty}^{\infty} Y_L (v,v) \, dv = 2 \int_{0}^{\infty}
\hat{Y}_L (k , - k ) \, dk \; ,
\end{equation}
which, when combined with Eq.\,(32), prove the difference inequality
Eq.\,(15) in this Schwartz-space $Y_L (v,v')$ case [in which the
integrand $(\langle T_{ab} \rangle_L -D_{ab})k^a k^b$ is integrable].
In the general case, the argument I need to use is again identical to
the one in [4] between Eqs.\,(4.\,40) and (4.\,59), except that in the
present analysis it leads to equalities of the form
\begin{eqnarray}
{{\mu}_{0L}}^M [ F_{\lambda , \kappa } , F_{\lambda , \kappa } ] & = &
\beta(\lambda , \kappa ) \left[
\mbox{$1\over 2$} \lambda \kappa + {\epsilon}_{1} ( \lambda \kappa )
\right] \; \nonumber \\
{{\mu}_{0L}}^M [ G_{\lambda , \kappa } , G_{\lambda , \kappa } ] & = &
\beta(\lambda , \kappa) \left[
\mbox{$1 \over 2$}
\lambda \kappa + {\epsilon}_{2} ( \lambda \kappa ) \right]
\; \\
{\sigma_L}^M [ F_{\lambda , \kappa } , G_{\lambda , \kappa } ]
& = & \beta(\lambda , \kappa ) \left[
\lambda \kappa + {\epsilon}_{3} ( \lambda \kappa ) \right] \;
\nonumber
\end{eqnarray}
with the same choices for $F_{\lambda , \kappa}(v)$ and $G_{\lambda ,
\kappa}(v)$ as in Eqs.\,(4.\,42). Here $\beta (\kappa , \lambda )$ is a
continuous function such that
\begin{equation}
0 \, < \, \beta(\lambda , \kappa ) \, < \, 1 \; \; \; \;
\; \; \; \; \forall \; \lambda > 0 , \; \kappa \geq 0 \;,
\end{equation}
and $\epsilon_{l}(x)$ are continuous functions with the same decay
property as described in [4] following Eq.\,(4.\,46). The second part of
the argument [spelled out in [4] between Eqs.\,(4.\,48) and (4.\,59)]
can be repeated identically here, leading to the inequality
\begin{equation}
\int_{- \infty}^{\infty} (\, \langle T_{vv} \rangle_L
\, - \, D_{vv} \, ) [ c(v/ \lambda )]^{2} \, dv
\, \geq \, \frac{1}{\lambda} \int_{0}^{\infty}
\beta (\lambda , y/\lambda ) \, \epsilon_{10}(y) \, dy \; .
\end{equation}
By Eq.\,(40) and the asymptotic behavior of
$\epsilon_{l}(x)$, there exist $\delta , \; K > 0$ such that
\begin{equation}
\int_{0}^{\infty} \beta ( \lambda , y/\lambda ) \, \epsilon_{10}
(y) \, dy \, \geq \,
- \int_{0}^{\infty} \frac{K}{(1+y)^{1+\delta}} \, dy
\, \equiv \, -C \; , \; \; \; \; \; \; C > 0 \; ,
\end{equation}
and, when combined with Eq.\,(41), this inequality proves not only the
difference-inequality result
\begin{equation}
\liminf_{\lambda \rightarrow \infty }
\int_{- \infty}^{\infty} (\, \langle T_{vv} \rangle_L
\, - \, D_{vv} \, ) [ c(v/ \lambda )]^{2} \, dv
\, \geq \, 0 \; ,
\end{equation}
but also the sharper estimate
\begin{equation}
\int_{- \infty}^{\infty} (\, \langle T_{vv} \rangle_L
\, - \, D_{vv} \, ) [ c(v/ \lambda )]^{2} \, dv
\, \geq \, - \frac{C}{\lambda}
\end{equation}
as in Eq.\,(4.59).

{}~~~~~~~

{\bf \noindent 5. Proof of the difference inequality in curved
two-dimensional spacetime}

Consider a two-dimensional, globally hyperbolic spacetime $(\SM ,g)$
with a massless Klein-Gordon field $\phi$, and
let $\gamma$ be a complete null geodesic in $\SM$. By Sect.\,5
of [4], if $\gamma$ is achronal
the renormalized stress-energy tensor in every Hadamard
state of $\phi$ satisfies the difference inequality Eq.\,(15) along
$\gamma$ with $D_{ab}=0$ (i.e., it satisfies
ANEC). If $\gamma$ is not achronal,
then I claim that $(\SM ,g)$ has topology
$S^1 \times \BR$ and is globally conformal to a flat cylinder $\BL$
for some $L>0$.

To prove this, let $p$ and $q$ be any pair of timelike-related
points along $\gamma$ (such pairs exist since $\gamma$ is assumed
non-achronal). Assume that $q \in I^{+}(p)$.
Since $(\SM ,g)$ is globally hyperbolic, it is causally
simple ([22]), i.e., $\dot{J}^{+}(p)$ consists of null geodesics from
$p$ which have no past endpoints other than $p$ itself.
Clearly, $q$ cannot belong to $\dot{J}^{+}(p)$ as it is
timelike-related to $p$; therefore, $\gamma$ must leave $\dot{J}^{+}(p)$
at some point $p'$ to the past of $q$. Since $\dot{J}^{+}(p)$ is an
achronal $C^{1-}$ submanifold without boundary
([22], Chapter 6), and $p'$ lies on
$\dot{J}^{+}(p)$, $\gamma$ must intersect the other null generator
$\delta$ of $\dot{J}^{+}(p)$ at $p'$;
otherwise $p'$ would lie on the boundary of
$\dot{J}^{(+)}(p)$. The portion of $\delta$
to the future of $p'$ is timelike related to $p$ (since every
point in this portion lies on a broken null geodesic from $p$),
therefore $\delta$ must also leave $\dot{J}^{+}(p)$ at $p'$; otherwise
$\dot{J}^{+}(p)$ would not be achronal.
Consequently, $\dot{J}^{+}(p)$ is compact. Because $(\SM ,g)$ is globally
hyperbolic, there exists a Cauchy surface $\Sigma$ through $p$, and any
global timelike vector field on $\SM$ provides a diffeomorphism
from $\dot{J}^{+}(p)$ into $\Sigma$. Since $\dot{J}^{+}(p)$ is compact
without boundary, $\Sigma$ must also be compact. But the only compact
1-manifold is $S^1$, and by global hyperbolicity $\SM$ is diffeomorphic
to $\Sigma \times \BR$, therefore, $\SM$ is diffeomorphic
to $S^1 \times \BR$. To prove that $(\SM ,g)$ is globally conformal to
$\BL$, it suffices to
simply carry out the usual local argument which proves
that any two-dimensional spacetime is locally conformally flat, using as
the time coordinate $t$ a smooth labeling of the Cauchy surfaces $S^1
\subset \SM$, and as the $x$ coordinate any coordinate on one of the
Cauchy $S^1$'s extended globally
onto $\SM$ (apart from the obvious coordinate
singularity on $S^1$) by keeping it constant along a timelike vector
field orthogonal to the Cauchy surfaces.
To summarize: if a globally hyperbolic
$(\SM, g)$ admits a non-achronal null geodesic, then
there exists a diffeomorphism $\Psi : \BL \rightarrow \SM$ such that
${\Psi}^{\ast} g = C(u,v) {\eta}_L$,
where $\eta_L=-du\, dv$ is the flat metric
on $\BL$ written in local null coordinates $\{u,v\}$, and $C(u,v) >0$
is a smooth function on $\BL$.

With $\gamma$ a complete
non-achronal null geodesic in $(\SM ,g)$, and with the
simple geometry of $\SM$ as uncovered in the above paragraph, it is now
quite straightforward to give an analysis parallel to that of Sect.\,5
in [4], where the proof of ANEC in curved two-dimensional spacetime was
reduced to the preceding proof in flat Minkowski space. Namely,
the Casimir vacuum state on $\BL$ given by Eq.\,(28) determines, under
the conformal isometry $\Psi : \BL \rightarrow \SM$, a corresponding
quantum state on $\SM$. [This is because the massless wave operator
as well as the Klein-Gordon inner product $\sigma$ are
conformally invariant in two dimensions, consequently there
exists a one-to-one correspondence between states defined on
$\BL$ and states defined on $\SM$, determined
by mapping the two-point distributions backwards or forwards
via the diffeomorphism $\Psi$. Note also that the spacetime
$(\SM ,g)$ is in fact {\sl isometric} to the cylinder $\BL$ equipped with
the metric $g = C \, \eta_L =-C(u,v) \, du \, dv$, therefore
I can use this representation of $(\SM ,g)$
throughout without any loss of generality,
and as this will simplify my notation considerably
I will do so from here on.] Let me denote the two-point
distribution of this state by $\mu_c (x,x')$ [in the isometric
representation of $\SM$ as $(\BL ,g)$, this distribution has the
same functional form as ${\mu}_{0L}(x,x')$].
The renormalized stress-energy in this
``conformal" Casimir vacuum is determined entirely by the conformal
anomaly [see Eqs.\,(10)--(11) in [3]], and can be written in the form
\begin{equation}
D_{ab} \, = \, \frac{1}{C} \, D^{(0)}_{ab}
\, + \,
\frac{1}{48 \pi} R g_{ab} \, + \, {\theta}_{ab}
\; ,
\end{equation}
where $D^{(0)}_{ab}$ is the Casimir energy on the flat $\BL$ given
by the right hand side of Eq.\,(14), $R$ is the Ricci scalar of
$(\SM ,g)$, and, in the local null
coordinates $\{u,v\}$,
\begin{eqnarray}
{\theta}_{uu} & = & \frac{1}{24 \pi} \left[ \frac{C_{,uu}}{C}
\, - \, \frac{3}{2} \frac{{C_{,u}}^{2}}{C^{2}} \right] \; , \nonumber \\
{\theta}_{vv} & = & \frac{1}{24 \pi} \left[ \frac{C_{,vv}}{C}
\, - \, \frac{3}{2} \frac{{C_{,v}}^{2}}{C^{2}} \right] \; , \\
{\theta}_{uv} & = & {\theta}_{vu} = 0 \; . \nonumber
\end{eqnarray}
Now, any Hadamard state on $(\SM ,g)$ has a two-point function of the
form
\begin{equation}
\mu (x,x') = \mu_c (x,x') + w(x,x') \; ,
\end{equation}
where $w(x,x')$ is a smooth bi-solution such that $\mu$ satisfies the
positivity inequality Eq.\,(19). It is crucial to keep in mind that
although $\mu_c (x,x')$ has the Hadamard form, it is {\sl not} a locally
constructed two-point distribution, hence it cannot be used to
regularize $\mu (x,x')$ as Eq.\,(47) suggests. Instead,
an appropriate Hadamard distribution $\mu_H$ constructed entirely out
of the local geometry of $(\SM ,g)$ needs to be subtracted
from $\mu$ to obtain the regularized two-point function; the components
of the stress-energy tensor are then obtained as the
coincidence limits of the derivatives of this regularized two-point
function $\mu -\mu_H$. [This point is of course
also valid for the analysis of the previous section, where it was
implicit in the derivation of Eq.\,(32) from Eq.\,(31).] However, for my
analysis here (as also for the analysis of the
previous section), it is not necessary to make
explicit the form of $\mu_H (x,x')$; the stress-energy due to the
difference $\mu_c - \mu_H$ is already determined completely by the
Casimir contribution Eqs.\,(45)--(46), and the rest of the stress-energy
is given simply by coincidence limits of the appropriate derivatives
of the smooth bi-solution $w(x,x')$. Therefore, expressions of the form
Eqs.\,(31)--(32) (with $w$ replacing $w_L$)
are still valid for the difference-inequality integrand
$( \langle T_{ab} \rangle - D_{ab} )k^a k^b$ along $\gamma$, and using
the positivity inequality for the two-point function Eq.\,(47) in
exactly the same manner as I did in the previous section, I arrive at
the following final \vspace{5pt}conclusion:
\newline{\bf Theorem}: Let $(\SM ,g)$ be a globally
hyperbolic two-dimensional spacetime with a massless
Klein-Gordon field $\phi$. Then the regularized
stress-energy tensor
$\langle \omega | T_{ab} | \omega \rangle$ of $\phi$
in any Hadamard state $\omega$ is constrained in
the following way: (i) Along every {\sl achronal} complete null geodesic
$\gamma \subset \SM$ the difference inequality Eq.\,(15) holds with
$D_{ab}=0$. (ii) If $\SM$ admits a non-achronal null geodesic, then
$(\SM ,g)$ is globally conformal
to $\BL \cong S^1 \times \BR$, and along every
complete null geodesic $\gamma$ in $\SM$ the difference inequality
Eq.\,(15) holds with $D_{ab}$ defined by Eqs.\,(45)--(46).

The proof of the difference inequality in two dimensions suggests that
more generally,
when $(\SM ,g)$ is a multiply-connected
(four-dimensional) globally hyperbolic
spacetime, an inequality of the form Eq.\,(15) might hold on a complete
non-achronal null geodesic $\gamma$ if the lifting
of $\gamma$ in the simply-connected
covering space $\tilde{\SM}$ is achronal and satisfies ANEC. When
$\gamma$ is a non-achronal complete null geodesic in a {\sl
simply-connected} spacetime
(i.e., when the failure of achronality is due to gravitational focusing
rather than the topology of $\SM$),
my proof does not provide any insights into
whether difference inequalities are reasonable
as constraints on the stress-energy tensor along $\gamma$.

{}~~~~~~~

{\bf \noindent 5. Acknowledgements}

Conversations with L.\ Ford and T.\ Roman at the seventh
Marcel Grossman meeting motivated this work; I am grateful to them for
their insights and encouragement. I am also
grateful to R.\ Wald for providing very useful information
via e-mail. This research was carried out at the Jet
Propulsion Laboratory, Caltech, and was
sponsored by the NASA Relativity Office and by the National Research
Council through an agreement with the National Aeronautics and Space
Administration.

\newpage

\begin{center}
{\bf REFERENCES}
\end{center}

\noindent{\bf 1.} R.\ Wald, private communication. That the pointwise
quantity $\langle \omega | \!\! : \!\! T_{00}(x) \!\! : \!\!
| \omega \rangle$ is unbounded from below is easy to see from the
following argument pointed out to me by R.\ Wald: Once we find
any state $\omega$ for which $\langle
\omega | \!\! : \!\! T_{00}(x) \!\! : \!\!
| \omega \rangle$ is negative, for example by the argument described in
the introductory paragraph of [3], we can modify
$\omega$ by
scaling the frequency of each of its
particle modes uniformly up while keeping its particle content
intact; this will scale
the quantity $\langle \omega | \!\! : \!\! T_{00}(x) \!\! : \!\!
| \omega \rangle$ by the same factor, and make it unboundedly large and
negative as the scaling factor goes to infinity. The results about the
volume integral of $\langle \omega | \!\! : \!\! T_{00}(x) \!\! : \!\!
| \omega \rangle$ over finite boxes were obtained by D.\ Garfinkle while
he was a graduate student at the University of Chicago and remain
unpublished. Note that by Wald's argument,
to guarantee the unboundedness from below of the integral of
$\langle \omega | \!\! : \!\!
T_{00}(x) \!\! : \!\!
| \omega \rangle$ over a given finite box in spacetime
it suffices to find a single state $\omega$ for which this integral is
negative.

\noindent{\bf 2.} G.\ Klinkhammer, Phys.\ Rev.\ D {\bf 43}, 2542 (1991).

\noindent{\bf 3.} U.\ Yurtsever, Class.\ Quantum Grav.\ {\bf 7}, L251
(1990).

\noindent{\bf 4.} R.\ Wald and U.\ Yurtsever, Phys.\ Rev.\ D {\bf 44},
403 (1991).

\noindent{\bf 5.} L.\ H.\ Ford and T.\ A.\ Roman, Phys.\ Rev.\ D {\bf 46},
1328 (1992); {\bf 48}, 776 (1993).

\noindent{\bf 6.} G.\ Klinkhammer, Caltech Preprint GRP-321 (1992).

\noindent{\bf 7.} A.\ Folacci, Phys.\ Rev.\ D {\bf 46}, 2726 (1992).

\noindent{\bf 8.} E.\ Flanagan, in preparation.

\noindent{\bf 9.} L.\ H.\ Ford and T.\ A.\ Roman, Phys.\ Rev.\ D {\bf 41},
3662 (1990).

\noindent{\bf 10.} F.\ J.\ Tipler, Phys.\ Rev.\ D {\bf 17}, 2521 (1978).

\noindent{\bf 11.} T.\ A.\ Roman, Phys.\ Rev.\ D {\bf 33}, 3526 (1986);
{\bf 37}, 546 (1988).

\noindent{\bf 12.} A.\ Borde, Class.\ Quantum Grav.\ {\bf 4}, 343 (1987).

\noindent{\bf 13.} R.\ Penrose, R.\ Sorkin, and E.\ Voolgar, Syracuse
University Preprint (1994).

\noindent{\bf 14.} That the averaged weak energy condition
fails to hold in general
along timelike geodesics in curved two-dimensional spacetime can
be seen from the analysis in [3]. Another, more interesting
counterexample is provided by four-dimensional
Minkowski spacetime in the presence
of conducting boundaries (i.e., boundaries on which the fields are
constrained to vanish). When the conductors are stationary (i.e., have
time-independent positions in spacetime), the stress-energy
tensor for a {\sl conformally coupled}
scalar field in the static vacuum state has been calculated by
Deutsch and Candelas in [15]. The stress-energy has a contribution that
diverges as one approaches a conductor;
this contribution is proportional
to the extrinsic curvature of the conducting boundary, where the extrinsic
curvature is calculated with respect to the outward-pointing normal,
and consequently has a positive sign when the boundary is convex, and a
negative sign when it is concave. If the conducting
surface contains a flat direction,
the divergent contribution projected on that direction vanishes
(due to conformal invariance). It is then clear that
along any static, complete timelike geodesic that lies
at rest sufficiently close
to a {\sl concave} conductor the averaged weak energy condition will be
violated. Because of the geometry,
no complete {\sl null} geodesic can ever come arbitrarily close to a
concave conductor. A complete null geodesic can graze arbitrarily
closely by a {\sl convex} boundary, or a boundary which contains
a flat direction (and then only
along that direction); but in these cases the divergent contribution
from the stress-energy is nonnegative.
In fact, ANEC holds; in the static, conformal vacuum
state, in which the regularized two-point
function $w(x,x')$ satisfies appropriate asymptotic fall-off conditions,
the analysis given in [4]---modified for conformal coupling---is
sufficient to prove ANEC along complete
null geodesics in Minkowski spacetime
in the presence of (stationary) conducting boundaries. The idea of using
conducting boundaries in flat spacetime as a testing ground for averaged
energy conditions is originally due to Curt Cutler.

\noindent{\bf 15.} D.\ Deutsch and P.\ Candelas, Phys.\ Rev.\ D {\bf 20},
3063 (1979).

\noindent{\bf 16.} M.\ Visser, Washington University Preprint (1994).

\noindent{\bf 17.} U.\ Yurtsever, ``A note on the averaged null energy
condition in quantum field theory" (in preparation).

\noindent{\bf 18.} L.\ H.\ Ford and T.\ A.\ Roman, Tufts University
Preprint (1994).

\noindent{\bf 19.} B.\ S.\ Kay and R.\ Wald, Phys.\ Rep.\ {\bf 207},
49 (1991).

\noindent{\bf 20.} G.\ Gonnella and B.\ S.\ Kay, Class.\ Quantum Grav.\
{\bf 6}, 1445 (1989).

\noindent{\bf 21.} Y.\ Katznelson, {\it An Introduction to Harmonic
Analysis} (Dover Publications, New York 1976).

\noindent{\bf 22.} S.\ W.\ Hawking and G.\ F.\ R.\ Ellis,
{\it The Large Scale
Structure of Space-time} (Cambridge University Press, Cambridge 1973).

\newpage

\begin{center}
{\bf FIGURE CAPTIONS}
\end{center}

\noindent{\bf Figure 1.} The geometry involved in the analysis leading
to Eq.\,(5). The spacelike surfaces $S_{\alpha}$ are the level sets of
the time function $\alpha$; accordingly, $S_{0}$ is $S$ and $S_{1}$ is
the future horizon $H^{+}(S)$.

\end{document}